\documentclass[aps,prl,showpacs,twocolumn]{revtex4}
\usepackage{amsmath,amssymb,amsthm,amscd,amsfonts}
\usepackage{graphicx}
\usepackage{colordvi}
\usepackage{color}

\def\be{\begin{equation}}
\def\ee{\end{equation}}
\def\bee{\begin{eqnarray}}
\def\ene{\end{eqnarray}}
\def\bes{\begin{subequations}}
\def\ees{\end{subequations}}

\begin{document}

\title{Solitons in a medium with linear dissipation and localized gain}

\author{Dmitry A. Zezyulin,$^{1,*}$ Yaroslav V.  Kartashov,$^2$ and Vladimir V. Konotop$^{1}$ }

\affiliation{ $^1$Centro de F\'{\i}sica Te\'orica e Computacional, and Departamento de F\'{\i}sica, Faculdade de Ci\^encias,  Universidade de Lisboa,
Avenida Professor Gama Pinto 2, Lisboa 1649-003, Portugal
\\
$^2$ICFO-Institut de Ciencies Fotoniques, and Universitat Politecnica de Catalunya, Mediterranean Technology Park, 08860 Castelldefels (Barcelona), Spain
}

\date{\today}

\begin{abstract}
We present a variety of dissipative solitons and breathing modes in a medium with  localized gain and homogeneous linear dissipation. The system posses a number of unusual properties, like exponentially localized modes in both focusing and defocusing media, existence of modes in  focusing media at negative propagation constant values, simultaneous existence of stable symmetric and anti-symmetric localized modes when the gain landscape possesses two local maxima, as well as the existence of stable breathing solutions.
\end{abstract}

\pacs{42.65.Tg, 42.65.Sf}
\maketitle

\noindent

Dissipative solitons appear in a variety of forms and settings. They form in laser systems with saturable gain and absorption~\cite{Rosanov}, in homogeneous materials where light evolution can be described by the cubic-quintic Ginzburg-Landau (GL) equation~\cite{Akhmediev}, and in systems with spatially localized gain and uniform nonlinear
losses~\cite{Malomed,KKVT,Kutz,KKVT1,KKV,Malomed2}. A characteristic feature of all these  systems is that the existence of dissipative solitons  there is mediated by the competition of gain and losses having different physical origin. Thus, in the cubic-quintic GL equation solitons exist because of exact balance of the linear losses, cubic gain, and quintic losses. Likewise, solitons can also form because cubic losses compensate the spatially localized gain.  This has been established in one-dimensional settings  in the presence of a linear localized potential~\cite{Malomed} or    lattice~\cite{KKVT}, in two-dimensional problems in planar waveguide arrays~\cite{Kutz}, or lattices~\cite{KKVT1}, or even in two-dimensional systems without additional refractive index modulations ~\cite{KKV}.

It is therefore important to elucidate simpler physical settings, where only the linear gain and losses result in formation of dissipative solitons.
This is the goal of this Letter. More specifically, we report on the existence of stable dissipative solitons supported by a localized gain in  a homogeneous Kerr  medium where only the linear losses are present far from the center of gain landscape. This simple system possesses a number of interesting properties
among which  is the existence of localized modes even for negative propagation constant values, simultaneous stability of symmetric and antisymmetric  modes in two-humped gain landscapes, as well as the existence of breathing modes.

We consider the nonlinear Schr\"odinger (NLS) equation
\begin{equation}
\label{eq:main} iq_z = -q_{xx} -i[\gamma_0-\gamma(x)] q + \sigma
|q|^2q.
\end{equation}
where   $\sigma = 1$ ($\sigma = -1$) corresponds to defocusing (focusing) nonlinearity,   $\gamma(x)$  describes a localized gain  concentrated in the domain (or domains) having a characteristic width $d$.
Without loss of generality we rescale the background dissipation to the value:  $\gamma_0=1$, and consider the situation where $\max \gamma(x)>1$.

We look for spatially localized solutions obeying  $\lim_{|x|\to \infty}|q(z,x)| = 0$ and start with the stationary modes $q(z, x) = e^{ibz} w(x)$
where $w(x) = e^{i\theta(x)}u(x)$ with $u(x)\geq 0$ and $\theta(x)$ being real functions. Then, introducing the current density
$j(x)  \equiv  \theta_xu^2$
one obtains from (\ref{eq:main}) the system describing stationary modes
\begin{eqnarray}
\label{hydro1}
u_{xx} -bu+\sigma u^3- \frac{j^2}{u^3}=0,
\quad
j_x+[1-\gamma(x)]u^2=0
\end{eqnarray}
In order to establish the domain of possible variation of the propagation constant $b$, we now address the limit $x\to\infty$ focusing on the exponentially decaying solutions, i.e. assuming $\displaystyle{u\sim Ce^{-\mu x}}$, where $C$ and $\mu$  are positive constants. From the second of Eqs. (\ref{hydro1}) it readily follows that in this limit $\displaystyle{j\sim  C^2   e^{-2\mu x}/2\mu}$. This suggests that in the first of equations (\ref{hydro1})  the term $\sigma u^3$ can be neglected in the limit $x\to\infty$, while substituting the asymptotic values of $u$ and $j$ in the terms which are left we obtain  $b=\mu^2-1/(4\mu^2)$.
It follows from this formula  that the propagation constant for localized solutions can be either positive or negative even if  the nonlinearity is focusing (contrary to what happens in uniform conservative systems or in dissipative systems~\cite{Malomed,KKVT,KKVT1,KKV}). If $b\to+ \infty$ then $\mu\to +\infty$ while if $b\to -\infty$ then  $\mu\to +0$.   Respectively, there must exist localized modes with $\mu>0$ even at $b=0$ while the spatial delocalization occurs in the limit $b\to -\infty$.

The condition of balance between the dissipation and gain (for $\gamma_0=1$) has the form $U=\int\gamma(x)|u|^2dx$ where $U=\int |u|^2\ dx$ is the energy flow. From the balance between the nonlinearity and diffraction we obtain
\begin{eqnarray}
\label{energy}
 b U=-\int |w_x|^2dx-\sigma\int |w|^4dx
\end{eqnarray}
From this formula one readily concludes that for zero propagation constant, $b=0$, the mode should decay with $\mu= 2^{-1/2}$, i.e. $|w_x|^2\not\equiv 0$, and hence (\ref{energy}) can be satisfied only in the focusing medium (i.e. at $\sigma = -1$). Thus the linear limit  of the problem (i.e. $U\to 0$) corresponds to a localized mode with $b=b^{(0)}<0$ (indeed, in this case
$\int |w|^4dx/U\leq \max{|w|^2}\to 0$ and can be neglected in (\ref{energy})). In the defocusing medium
exponentially localized solutions exist only if $b<b^{(0)}$ while
in the  focusing medium dissipative solitons can exist with both positive and negative  propagation constants, i.e. at $b>b^{(0)}$.

\begin{figure}[htb]
\vspace{-1.5cm}
\centerline{
        \includegraphics[width=\columnwidth]{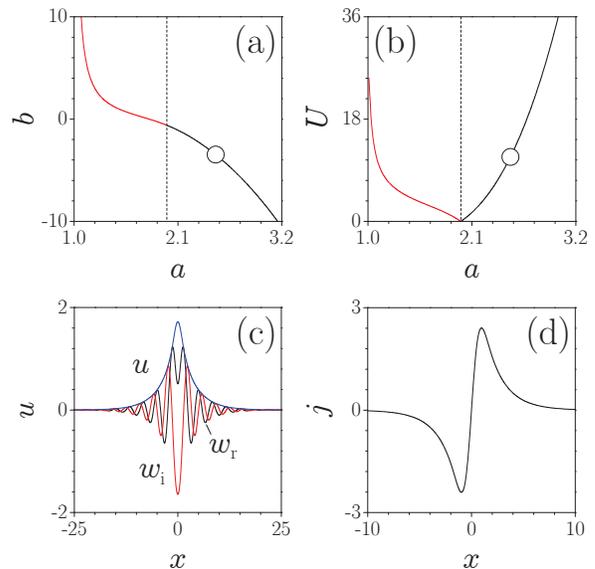}}
\vspace{-3.2cm}
\caption{Propagation constant (a) and energy flow (b) {\it vs}  $a$ for solitons supported by single gain channel. Red and black curves correspond to focusing and defocusing media.  Dashed line indicates  $a$ value corresponding to the linear limit. Profile of soliton (c) and current density (d) in defocusing medium at $a=2.5$   corresponding to the circles in (a) and (b).}
        \label{fig-fam0}
        \vspace{-0.3cm}
    \end{figure}

We start with the case of the gain  localized  in one spatial domain, setting  $\gamma(x)=   ae^{-x^2/d^2}$, ($a > 1$ characterizes the gain strength).
By fixing the width of the gain landscape $d$ and changing $a$ we obtain
the dependencies $b(a)$ and $U(a)$, which are depicted in Fig.~\ref{fig-fam0} for  $d=1$. These dependencies  bifurcate from the linear limit corresponding to $U=0$ and  to  $ b^{(0)}\approx -0.61$. 
{The very fact of existence of localized linear modes with $b=b^{(0)}$ in inhomogeneous gain landscapes is known as the gain-guiding effect (see e.g.~\cite{Siegman}).
In the linear case the guiding occurs for a particular value of $a=a^{(0)} \approx  1.98$.   Focusing nonlinearity diminishes gain at which localization occurs, while in a defocusing medium  a higher gain is required for localization (Fig.~\ref{fig-fam0}).}
In Fig.~\ref{fig-fam0} we also show
an example of a nonlinear mode.
Increase of the gain results in decrease of $U$ in the focusing medium, that is also accompanied by expansion of the soliton. In the defocusing medium the growth of $a$ results in increase of the energy flow and progressive expansion of the soliton outwards  the gain domain. Solitons are exponentially localized in the both focusing and defocusing media, {as predicted by
the analysis of (\ref{hydro1})  which showed that the current results in effective renormalization of the propagation constant. Physical understanding of the exponential localization in the defocusing medium, comes also form the observation that the current results in the effective self-focusing of the beam, what stems form the negative sign in front of $j^2/u^3$.}

\small
\normalsize

\begin{figure}[htb]
\vspace{-3.8cm}
\centerline{
        \includegraphics[width=\columnwidth]{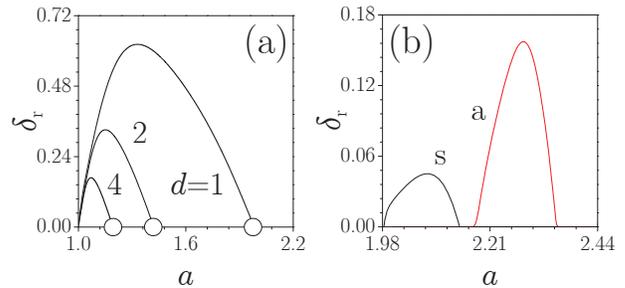}%
        }
\vspace{-4.1cm}
\caption{
{(a) Instability increment $\delta_r$ {\it vs} $a$
 for solitons in a single gain channel in focusing medium. Circles indicate points beyond which solitons exist only in defocusing medium where they are stable. (b) $\delta_r (a)$ for symmetric "s" and antisymmetric "a" solitons in two gain channels with $d=1$ in defocusing medium.}
}
\vspace{-0.3cm}
\label{fig:stability}
\end{figure}

{Linear stability analysis (confirmed by the direct propagation)} indicates that the solitons in the focusing medium are unstable [see Fig.~\ref{fig:stability}(a)],
 while  in defocusing medium they are attractors and can be excited with a variety of inputs ranging from noisy to localized patterns. {This  difference in the stability can be understood from
 Eqs.~(\ref{hydro1}). The solution represents a flow outwards the "source" (i.e. the gain domain). The defocusing nonlinearity enhances the outflow from the high intensity region  thus contributing to the stability, while the focusing medium enhances the field concentration in the gain domain, what stimulates further growth of the peak amplitude.}
\small
\normalsize

\begin{figure}[t]
\vspace{-.5cm}
\centerline{
        \includegraphics[width=\columnwidth]{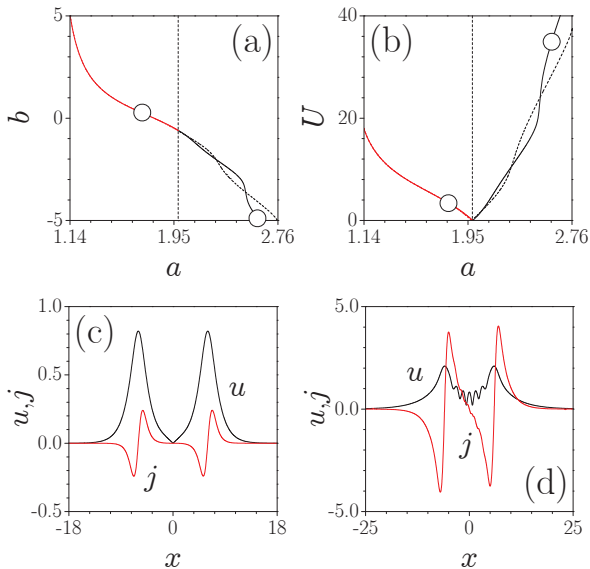}
}
\vspace{-3.2cm}
\caption{Propagation constant (a) and energy flow (b) {\it vs} $a$   for solitons supported by the two gain channels at $x_0=6$. Solid and dashed lines correspond to the symmetric and antisymmetric modes. Red and black lines correspond to the focusing and defocusing media. The circles correspond to the solitons shown in (c),(d). (c) Unstable antisymmetric soliton in the focusing medium at  $a=1.8$. (d) The stable symmetric soliton in defocusing medium at   $a=2.6$.  }
        \label{fig:fam1}
        \vspace{-.5cm}
    \end{figure}

Next we  study  the case of the gain landscape with two maxima  at
  $x=\pm x_0$ considering
$
\gamma(x)=   a  e^{-(x-x_0)^2/d^2} + a e^{-(x+x_0)^2/d^2},
$
where $a  > 1$.
Here  one also can obtain modes both for focusing and defocusing media.
In   Fig.~\ref{fig:fam1} we present  symmetric  (even)  and antisymmetric   (odd) modes.
Both  types of solutions bifurcate from the linear limit. The values of parameters $a$ and $b$ at which bifurcations take place are very close ($b_s^{(0)}\approx -0.61$,  $a_s^{(0)}\approx 1.99$ for symmetric, and  $b_a^{(0)}\approx -0.60$,  $a_a^{(0)}\approx 1.98$ for antisymmetric solitons). Closely to the linear limit and in the focusing medium  the  $U(a)$ curves for the symmetric and antisymmetric solitons are almost indistinguishable, while in the defocusing medium their energy flows may differ considerably [Figs.~\ref{fig:fam1} (a) and (b)]. The examples of modes
are shown in Figs.~\ref{fig:fam1} (c) and (d).
\small
\normalsize

Linear stability analysis reveals 
multiple alternating  stability and instability domains (in $a$)  for both symmetric and antisymmetric modes [Fig.~\ref{fig:stability} (b)].
A remarkable fact is that stable symmetric and antisymmetric modes co-exist at the same parameters of the system.
Moreover,  for the same value of $a$ both symmetric and antisymmetric modes in the defocusing medium can be attractors.

\small
\begin{figure}
        \includegraphics[width=\columnwidth]{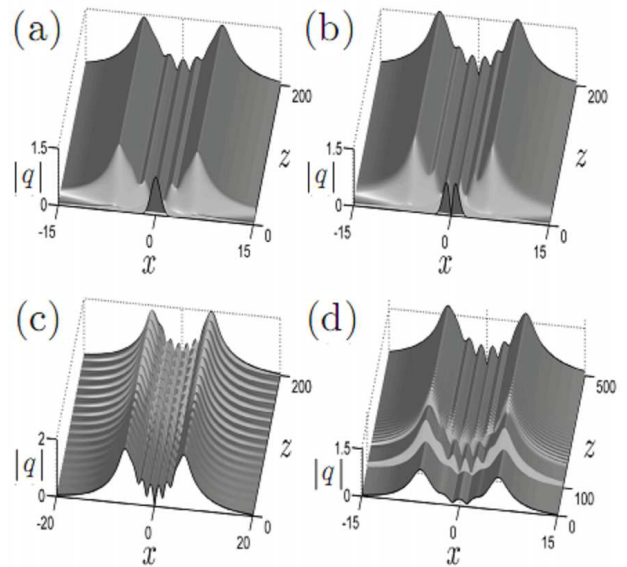}
\caption{ Excitation of stable symmetric (a) and antisymmetric (b) solitons by symmetric and antisymmetric input beams in a system with two gain channels for $a=2.391$. (c) Excitation of a breather at $a=2.61$  starting with the unstable antisymmetric soliton. (d) Switching between stable symmetric and antisymmetric solitons stimulated by abrupt change of $a$   at  $z=100$.}
\label{fig-evolutions}
\vspace{-0.1cm}
\end{figure}
\normalsize

\begin{figure}[htb]
\centerline{
        \includegraphics[width=\columnwidth]{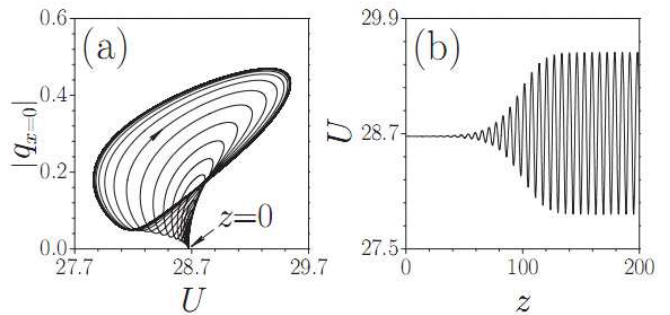}%
        }
\caption{The trajectory
(panel  a) and dependence $U(z)$ (panel  b) corresponding to the dynamics shown in Fig.~\ref{fig-evolutions}(c).
The bold line shows the limit cycle.
}
\label{fig_limcicle_proj}
\end{figure}

We found that  for certain values of $a$  even input beams evolve into the symmetric  stationary mode while   odd input beams evolve into the antisymmetric  mode, see panels (a) and (b) of Fig.~\ref{fig-evolutions}. When one of the modes is unstable it usually evolves into a stable mode or into
\textit{pulsating mode}. An example of such behavior
is shown in Fig.~\ref{fig-evolutions} (c) where  perturbed unstable antisymmetric mode transforms into the breather.
(We show only the initial stage of the evolution;  simulations were performed up to  $z=2\cdot10^3$). Dynamics shown in Fig.~\ref{fig-evolutions}(c) is also illustrated in Fig.~\ref{fig_limcicle_proj} where a corresponding trajectory in the phase space and dependence $U(z)$ are plotted.

\small
\normalsize

Finally we explore the possibility of switching between stable solutions by  changing the control parameter $a$.  An example
is shown in Fig.~\ref{fig-evolutions} (d) where the input beam corresponds to the stable symmetric mode at $a = 2.186$.  At  $z=100$ the strength of the gain $a$ is switched from $2.186$ to $2.5$. Since the symmetric mode at $a =2.5$ is unstable, the switching  results in the beam evolution into the stable antisymmetric mode corresponding to $a=2.5$.

To conclude, we found
a variety of unusual properties of the solitons described by model (\ref{eq:main}), such as existence of exponentially localized modes  with negative propagation constant values in both focusing and defocusing media, simultaneous stability of symmetric and antisymmetric solitons, and the existence of persistent breathers.

DAZ and VVK were supported
by the grants  SFRH/BPD/64835/2009
and PTDC/FIS/112624/2009.

\end{document}